\documentclass[aps,pra,reprint,a4paper]{revtex4-1}
\usepackage[utf8]{inputenc}
\usepackage{units}
\usepackage{graphicx}
\usepackage{longtable}

\begin{document}
 \title{Measurement of absolute transition frequencies of ${}^{87}\mathrm{Rb}$ to $n S$ and $n D$ Rydberg states by means of electromagnetically induced transparency }
 \date{\today}
 \author{Markus \surname{Mack}}
 \email{mack@pit.physik.uni-tuebingen.de}
 \author{Florian \surname{Karlewski}}
 \author{Helge \surname{Hattermann}}
 \author{Simone \surname{H\"ockh}}
 \author{Florian \surname{Jessen}}
 \author{Daniel \surname{Cano}}
 \author{J\'ozsef \surname{Fort\'agh}}
 \email{fortagh@uni-tuebingen.de}
 \affiliation{CQ Center for Collective Quantum Phenomena and their Applications, Physikalisches Institut, Eberhard-Karls-Universit\"at T\"ubingen, Auf der Morgenstelle 14, D-72076 T\"ubingen, Germany}
	      
 \begin{abstract}
	 We report the measurement of absolute excitation frequencies of ${}^{87}\mathrm{Rb}$ to $n S$ and $n D$ Rydberg states. The Rydberg transition frequencies are obtained by observing electromagnetically induced transparency on a rubidium vapor cell. The accuracy of the measurement of each state is $\lesssim \unit[1]{MHz}$, which is achieved by frequency stabilizing the two diode lasers employed for the spectroscopy to a frequency comb and a frequency comb calibrated wavelength meter, respectively. Based on the spectroscopic data we determine the quantum defects of ${}^{87}\text{Rb}$, and compare it with previous measurements on ${}^{85}\text{Rb}$. We determine the ionization frequency from the $5S_{1/2} (F\mathord=1)$ ground state of ${}^{87}\text{Rb}$ to $\unit[1 010.029 164 6(3)]{THz}$, providing the binding energy of the ground state with an accuracy improved by two orders of magnitude. 

 \end{abstract}
 \maketitle

\section{\label{sec:intro}Introduction}

Rydberg atoms are attracting large research interest as their quantum state can be prepared and controlled with high precision and flexibility by means of electromagnetic fields \cite{gallagher}. Actual research subjects include cavity quantum electrodynamics \cite{RevModPhys_CQED}, Rydberg-Rydberg interactions \cite{Comparat_Rydberg_Rydberg} with possible applications in quantum information processing \cite{RevModPhys_Rydberg_QI}, ultracold chemistry \cite{Rydberg_molecules} \cite{macrodimers} \cite{bendowski_nature}, and sensing dispersion forces between atoms and surfaces \cite{microcells} \cite{CP_metallic_surface} \cite{tauschinsky}. Due to their large polarizability, Rydberg atoms respond to static and dynamic electric fields with level shifts, and are therefore sensitive to interactions with microscopic and macroscopic objects. For precision measurements of such interactions and accurate quantum state control, it is beneficial to know the unperturbed Rydberg energy levels that we quantify in the present paper for the ${}^{87}\mathrm{Rb}$ atom, which is widely used in experiments.

In alkali atoms, the deviation from the hydrogen energy structure can be expressed in terms of quantum defects $\delta$ \cite{gallagher}. If the quantum defects are known, the energy of Rydberg levels can be calculated with respect to the ionization limit. The quantum defects of the $n S$ and $n D$ lines of ${}^{85}\mathrm{Rb}$ have been measured by Li et al. \cite{mmwave} using microwave excitation between Rydberg states. This method provides superior frequency resolution on the order of $\sim \unit[10]{kHz}$, however, it does not serve as an absolute frequency reference. If, in addition, absolute transition frequencies between the ground and Rydberg states are known, the binding energy of the ground state can be determined. 

Absolute measurements of the $nS$ and $nD$ lines of ${}^{85}\mathrm{Rb}$ as well as the isotope shift for ${}^{87}\mathrm{Rb}$ have been done by Stoicheff and Weinberger \cite{stoicheff_weinberger} using two-photon spectroscopy with an accuracy of $\unit[100]{MHz}$. Higher accuracy measurements include the $nF$ states of the ${}^{85}\mathrm{Rb}$ isotope with an accuracy of $\unit[8]{MHz}$ \cite{johnson_nF_2010} and on the $nS$ levels of ${}^{85}\mathrm{Rb}$ with an uncertainty greater than $\unit[6]{MHz}$ \cite{sansonetti}. In the present work, we measure Rydberg transition frequencies of ${}^{87}\mathrm{Rb}$ with an absolute accuracy of $\lesssim \unit[1]{MHz}$, which results from frequency stabilizing the lasers employed for the spectroscopy to absolute frequency references.

High resolution spectroscopy of Rydberg states is possible by observing electromagnetically induced transparency (EIT) \cite{Fleischhauer}. The detection of Rydberg states by means of EIT, and the measurement of the fine structure splitting of ${}^{85}\mathrm{Rb}$ have been reported by Mohapatra et al. \cite{adams}. In the present work, we implement EIT for measuring the absolute frequency of Rydberg excitations of ${}^{87}\mathrm{Rb}$, from which we determine the quantum defects of the $nS$ and $nD$ levels and the ground state ionization energy of this isotope.

\section{\label{sec:setup}EIT level scheme and experimental setup}

\begin{figure}[phtb]
 \includegraphics[width=0.25\textwidth]{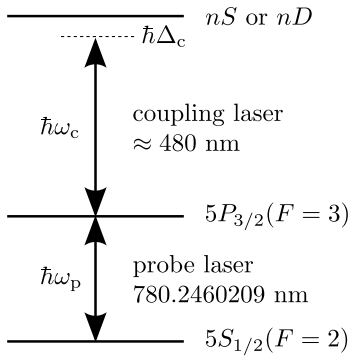}
 \caption{Energy level diagram of ${}^{87}\mathrm{Rb}$ and corresponding laser wavelengths for observation of electromagnetically induced transparency using a Rydberg state. The probe laser is stabilized to the lower atomic resonance, while the coupling laser, with a variable detuning $\hbar \Delta_\mathrm{c}$, is scanned across the Rydberg resonance.}
 \label{fig:excitation_scheme}
\end{figure}

The energy level diagram of ${}^{87}\mathrm{Rb}$ relevant for this work is shown in Fig.\,\ref{fig:excitation_scheme}. The three level system consists of the ${}^{87}\text{Rb}$ $5S_{1/2} (F\mathord=2)$ ground state, the $5P_{3/2} (F\mathord=3)$ intermediate state, and $nS$ or $nD$ Rydberg states with a high principal quantum number $n$. We observe EIT spectra on a rubidium vapor cell using two counterpropagating laser beams. The frequency of the ``probe laser'' is fixed at the $5S_{1/2} (F\mathord=2) \leftrightarrow 5P_{3/2} (F\mathord=3)$ transition, which is known to be $\unit[384.2281152]{THz}$ \cite{ye1996hyperfine} ($\mathrel{\hat{=}} \unit[780.2460209]{nm}$), while the ``coupling laser'' ($\unit[\approx 480]{nm}$) is scanned across the Rydberg resonance. We record EIT spectra by monitoring the probe laser transmission as a function of the coupling laser frequency. 

\begin{figure}[phtb]
 \includegraphics[width=0.45\textwidth]{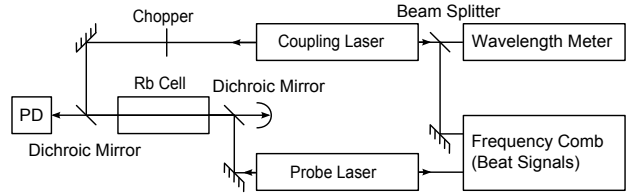}
 \caption{Optical setup (schematic) for the observation of EIT in a rubidium vapor cell. The probe and coupling beams are counterpropagating in the rubidium vapor cell (heated to $\sim \unit[45]{{}^\circ C}$) and the probe laser intensity is monitored with a photodiode (PD). A chopper in the coupling laser beam provides a reference signal for lock-in detection of the photodiode signal. The probe laser is frequency stabilized to a frequency comb. The coupling beam at $\unit[480]{nm}$ comes from a frequency doubled $\unit[960]{nm}$ laser that is stabilized and scanned by a wavelength meter. The wavelength meter is calibrated with the help of a frequency comb beat signal at $\unit[960]{nm}$.}
 \label{fig:setup}
\end{figure}

The experimental setup is illustrated in Fig.\,\ref{fig:setup}. Absolute frequency measurements are achieved by referencing the probe and coupling laser frequencies to a frequency comb (Menlo Systems ``FC 1500''). 

The source of the probe beam is a grating stabilized diode laser of $\sim \unit[500]{kHz}$ linewidth. The laser is stabilized to $\unit[780.2460209]{nm}$ by superimposing it with the beam of the frequency comb on a fast photodiode and locking the corresponding radio frequency beat signal with the nearest comb mode. The beat signal deviation, and thus the accuracy of the probe laser frequency, was maintained within $\unit[500]{kHz}$.

The coupling beam at $\unit[480]{nm}$ is sourced from a frequency doubled, grating stabilized diode laser (Toptica ``TA SHG Pro'') at $\unit[960]{nm}$ of $\sim \unit[500]{kHz}$ linewidth. In order to access any desired coupling frequency, we stabilize the $\unit[960]{nm}$ laser to a Fizeau interferometer based wavelength meter (HighFinesse ``WS Ultimate-2''). The wavelength meter with a built-in servo feedback unit measures frequency, and controls the frequency scans of the coupling laser. 

In order to assure the absolute accuracy of the wavelength meter, we perform an automated calibration procedure. A beat signal between the $\unit[960]{nm}$ laser and a known frequency comb mode is recorded with a digital spectrum analyzer. 
The beat frequency is determined by software peak recognition and the corresponding laser frequency is used to calibrate the wavelength meter. The resulting accuracy of frequency control by the calibrated wavelength meter was characterized with the frequency comb by exemplary measurements of the frequency offset with respect to the set frequency. We found the frequency offset after calibration approximately to be normally distributed with a standard deviation of $\unit[400]{kHz}$.

\begin{figure}[phtb]
\includegraphics[width=0.45\textwidth]{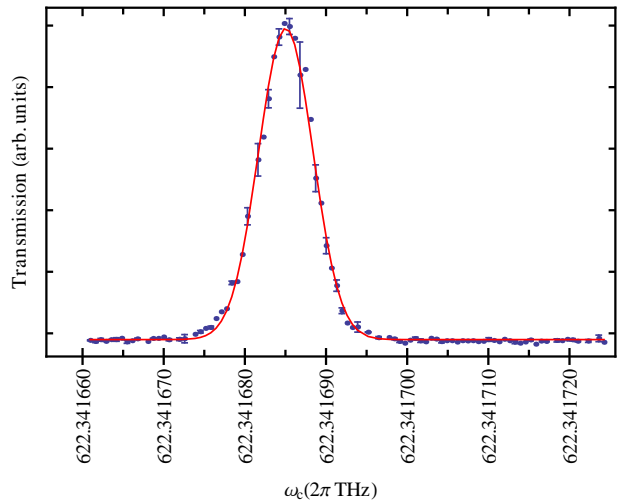}
\caption{(Color online) EIT resonance for the $34 S_{1/2}$ Rydberg state (exemplary) as recorded by the photodiode, against the coupling laser frequency measured simultaneously by the wavelength meter. A Gaussian shape is fitted for peak frequency determination (solid line).}
\label{fig:gnuplot_EIT_Zelle}
\end{figure}

The experimental sequence for recording EIT spectra on Rydberg levels is as follows. First, the wavelength meter is calibrated using the method described above. Second, a trigger is sent to the wavelength meter, that starts sweeping the coupling laser frequency (at most $\unit[12]{MHz/s}$ sweeping rate, $\unit[30]{MHz}$ span at $\sim \unit[960]{nm}$), while recording the measured coupling laser frequency. 
The photodiode signal of the probe beam is recorded by a digital storage oscilloscope. If the scan range covers an EIT resonance, a sharp transmission peak is recorded. In order to detect weak EIT signals of high $n$ states, the coupling laser beam is periodically blocked by a chopper wheel at a frequency of about $\unit[1]{kHz}$, and the photodiode signal of the probe beam is recorded using a lock-in amplifier synchronized to the chopper. As both the laser frequency and the photodiode signal are recorded simultaneously, we obtain the probe laser transmission as a function of the coupling laser frequency as shown in Fig.\,\ref{fig:gnuplot_EIT_Zelle}.
The measurement at each EIT resonance is fitted by a Gaussian function to determine the peak position.
In order to reduce the statistical error, the measurement sequence is performed five times for each Rydberg level.
		
\section{\label{sec:frequencies}Rydberg transition frequencies and measurement accuracy}

Absolute frequencies of transitions $5P_{3/2}(F\mathord=3) \to n S_{1/2} (F\mathord=2)$, $5 P_{3/2} (F\mathord=3) \to n D_{3/2}$ and $5 P_{3/2}(F\mathord=3) \to n D_{5/2}$ of ${}^{87}\mathrm{Rb}$ were observed for principal quantum numbers $n$ in the range $19-65$ ($nS$ states) and $19-57$ ($nD$ states). The results are listed in tables \ref{tab:E_n_table_s_1_2}, \ref{tab:E_n_table_d_3_2} and \ref{tab:E_n_table_d_5_2} (Appendix). We note that we were able to observe Rydberg states up to $n=180$ with the help of the lock-in amplifier. Such highly excited Rydberg states, however, exhibit large linewidths and partially irregular lineshapes, thus these are not analyzed in the present manuscript. For quantum defect calculations, we use only the data given in tables \ref{tab:E_n_table_s_1_2}, \ref{tab:E_n_table_d_3_2} and \ref{tab:E_n_table_d_5_2}.

The uncertainty of the frequency data comprise several known error sources of physical and technical nature. The main error source arises from the frequency stabilization of the probe laser with an uncertainty of $\unit[500]{kHz}$ (see section \ref{sec:setup}). A frequency deviation $\Delta_p$ from the  atomic resonance $\omega_p$ leads to the excitation of atoms with a velocity of $v = c \frac{\Delta_p}{\omega_p} $. If $\Delta_p$ is small compared to the Doppler broadening, the EIT transmission peak appears at a coupling laser frequency shifted by $\Delta_c = \frac{v}{c} \omega_c = (\omega_c/\omega_p) \Delta_p$ with respect to the resonance frequency $\omega_c$. Consequently, the uncertainty of the probe laser frequency translates into an uncertainty of $\unit[(\unit[780]{nm}/\unit[480]{nm}) \cdot 500]{kHz} \approx \unit[800]{kHz}$ of the EIT peak position measured with the coupling laser. 

The uncertainty of the coupling laser frequency enters the EIT peak determination with $\unit[200]{kHz}$. This is less than the uncertainty of the frequency stabilization of this laser ($\unit[400]{kHz}$) because of the five-fold repetition of the frequency measurement (including calibration). 

In our setup, the frequency of the comb modes can be determined with an absolute accuracy on the order of $\unit[10]{kHz}$. This uncertainty is negligible with respect to other sources of error.
The high accuracy of the frequency comb is achieved by a \unit[10]{MHz} GPS disciplined rubidium frequency reference (Precision Test Systems ``GPS10RBN'') with a specified Allan deviation of $2 \times 10^{-11}$ at $\unit[1]{s}$ observation time.
The accuracy of the transfer of the rf reference frequency to optical frequencies was determined by Kubina et al \cite{comb_comparison}, who compared two similar frequency combs referenced to one common rf frequency standard, finding the optical frequencies that were measured with both combs to agree to a level of $6\times 10^{-16}$.

The width of the observed EIT resonances is typically $\lesssim \unit[10]{MHz}$. The linewidth of the lasers, technical noise, triggering precision, and physical deviations from the Gaussian line shape that has been used for finding the peak position, contribute to the uncertainty of the measured resonance frequency as well. The overall uncertainty resulting from these effects was determined independently for each transition, and was typically less than $\unit[300]{kHz}$.
		
Despite the strong $n^7$ dependence of the polarizability \cite{gallagher}, line shifts due to stray electric fields are not present in the vapor cell due to screening by ions and electrons on the inner surface of the glass cell \cite{adams}. We verified the screening in our setup for field strengths up to $\unit[470]{V/cm}$ and on Rydberg states up to $n=150$. We observed line broadening due to external magnetic fields, which are in our setup on the order of the earth magnetic field, but no significant line shift or asymmetry of the resonance peak. Pressure shifts at a cell temperature of $\unit[\lesssim 45]{^\circ C}$ are calculated to be less than $\unit[200]{kHz}$ \cite{pressureshifts}.

 \section{\label{sec:rydberg}Quantum defects and ionization energy}
		
		The energies of Rydberg levels \cite{gallagher} are given by
		\begin{equation}
		 \label{eq:quantendefekt_formel}
	 		E_{n, l, j} = E_\mathrm{i} - \frac{\mathcal{R}^*}{(n - \delta(n, l, j))^2}\text{,}
	 \end{equation}
	 where $E_i$ is the ionization energy threshold and
	 \[
	 		\mathcal{R}^* = \frac{1}{1 + \frac{m_\mathrm{e}}{m_{{}^{87}\text{Rb}}}} \mathcal{R}_{\infty}
	 		= h \cdot \unit[3289.821 194 66(2)]{THz} 
	 \] 
	 is the Rydberg constant, corrected for the reduced electron mass in ${}^{87}\textrm{Rb}$.

  For sufficiently large principal quantum numbers $n$, the quantum defects $\delta(n, l, j)$ depend only little on $n$ and can be approximated by the modified Rydberg-Ritz parameters
		\begin{equation}
		 \label{eq:quantendefekt}
	 	 \delta(n,l,j) \approx \delta_0 + \frac{\delta_2}{(n - \delta_0)^2}\text{.}
	 	\end{equation}
		Within the accuracy of our measurements on $n \geq 19$ Rydberg states, the inclusion of higher order terms in the approximation did not lead to improved results.
		
		A precise measurement of the modified Rydberg-Ritz parameters for the $n S$, $n P$ and $n D$ series of ${}^{85}\mathrm{Rb}$ was performed by Li et al. \cite{mmwave}. By exciting microwave transitions between Rydberg levels ($n = 32$ up to $37$) and subsequent detection by field ionization, the transition frequencies were determined with a resolution of up to $\unit[10]{kHz}$. The ``average'' quantum defects $\delta(n, n + 1)$ determined by this measurement were used to improve the value of quantum defects previously obtained by Lorenzen and Niemax \cite{Lorenzen_Niemax}.

		Traditionally, the quantum defects are determined for the fine structure levels. The hyperfine splitting is negligible for $n D$ states, but not for $n S$ states because of a nonzero probability amplitude at the position of the nucleus. In our measurements of $n S$ states, only transitions to $F\mathord=2$ states were observed because of selection rules.
		In order to obtain quantum defects for the fine structure, we need to consider the hyperfine shifts in the evaluation of our data.
		
		The hyperfine splitting of states between $n=28$ and $33$ of both ${}^{85}\mathrm{Rb}$ and ${}^{87}\mathrm{Rb}$ was determined by Li et al \cite{mmwave}. Using their data, we extrapolate the hyperfine splitting between the $n S_{1/2} (F\mathord=1)$ and $(F\mathord=2)$ states of ${}^{87}\mathrm{Rb}$ according to a ${n^*}^{-3}$ law ($n^{*} := n - \delta(n,l,j)$) as
\[
 \Delta_{\mathrm{HFS}, F\mathord=2} - \Delta_{\mathrm{HFS}, F\mathord=1} = \unit[33.5(9)]{GHz} \cdot {n^*}^{-3} \text{.} 
\]

In general, the hyperfine shift is given by $\Delta_{\mathrm{HFS}, F} = \frac{A}{2} (F(F+1) - I(I+1) - J(J+1))$, which, compared with the measurement, yields the value of the hyperfine constant $A$. For the $J = 1/2$ states of ${}^{87}\mathrm{Rb}$ (nuclear spin $I=3/2$), the hyperfine splitting is $\Delta_{\mathrm{HFS}, F\mathord=2} - \Delta_{\mathrm{HFS}, F\mathord=1} = 2 A$.

Extending the extrapolation with the ${n^*}^{-3}$ law for the states $n \geq 28$ to $n = 19$ leads to a systematical error in the determination of $A$. In order to quantify this error we extrapolate further down to $n = 11$, for which the hyperfine splitting is known from a measurement by Farley et al. \cite{HFS_rubidium_cesium}. We find 10\% deviation between the calculated ($A_{11S_{1/2}} = \unit[34.5]{MHz}$) and the measured ($A_{11S_{1/2}} = \unit[37.4(3)]{MHz}$) value, that gives an upper limit of the systematic error for $n > 11$ states. 

We now can compute the hyperfine shift of the $n S_{1/2} (F\mathord=2)$ states that have been measured in this work,
\[
 \Delta_{\mathrm{HFS}, F\mathord=2} = \frac{3}{4} A = \frac{3}{2\cdot 4}\cdot \unit[34(3)]{GHz} \cdot {n^*}^{-3} \text{,}
\]
and subtract these frequency shifts from the measured transition frequencies to the $nS_{1/2} (F\mathord=2)$ states to find the quantum defects with respect to the fine structure levels. For the lowest $nS$ state measured in this work, $n=19$, the hyperfine shift is $\Delta_{\mathrm{HFS}, F\mathord=2} = \unit[3.2(3)]{MHz}$.

\begin{figure}[htpb]
 \includegraphics[width=0.45\textwidth]{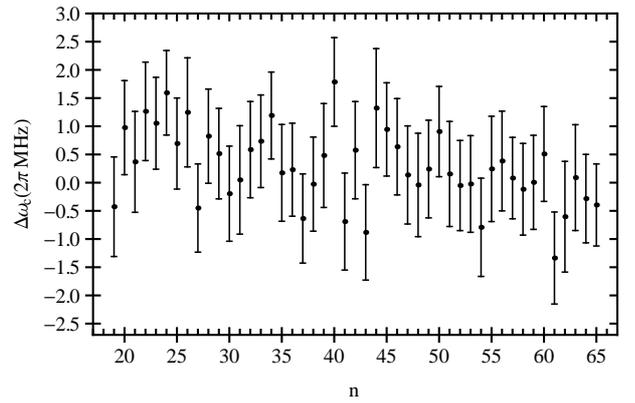}
\caption{Difference between measured transition frequencies to $n S_{1/2}$ Rydberg states (see table \ref{tab:E_n_table_s_1_2}) and the transition frequencies calculated with equation (\ref{eq:quantendefekt_formel}) based on the Rydberg-Ritz parameters and ionization frequency specified in tables \ref{tab:QD_Vergleich} and \ref{tab:IE_Vergleich})}
\label{fig:gnuplot_E_n_fit}
\end{figure}

In order to determine the modified Rydberg-Ritz parameters, equation (\ref{eq:quantendefekt_formel}) is simultaneously fitted to all three data series ($n S_{1/2}$, $n D_{3/2}$ and $n D_{5/2}$), with the ionization frequency $\omega_\mathrm{i} = E_\mathrm{i}/\hbar$ as a common fit parameter. The difference between measured frequencies and corresponding values obtained by the fit (residuals) are shown in Fig.\,\ref{fig:gnuplot_E_n_fit} exemplary for the measurement series on the $n S_{1/2}$ lines. The residuals vary around zero with an RMS deviation of $\lesssim \unit[0.75]{MHz}$, which applies to all three data series, consistent with the expected overall precision of the experimental setup.

The resulting values for the modified Rydberg-Ritz parameters are shown in table \ref{tab:QD_Vergleich} along the values determined by Li et al. \cite{mmwave} for ${}^{85}\mathrm{Rb}$. The parameters for the $nS$ series of ${}^{85}\mathrm{Rb}$ \cite{mmwave} and ${}^{87}\mathrm{Rb}$ (this work) match each other with comparable uncertainties. Omitting the hyperfine structure correction, the fit of our data returns a slightly different value of $\delta_0 = 3.131 178 8(8)$ (with no difference in $\delta_2$), that gives the value of the quantum defect of the $n S_{1/2} (F\mathord=2)$ hyperfine levels. 

The values of $\delta_0$ of the $nD$ series show a deviation from the values of Li et al.\ of about three times the stated uncertainty. 
Nonetheless, comparing the frequency differences measured in this work ($\omega_{c,n+1} - \omega_{c,n}$) with the Rydberg - Rydberg transition frequencies $\omega_{n, n+1}$ measured by Li et al., we find that the values available in both data sets are well within one standard deviation ($nS_{1/2}$ and $nD_{5/2}$) or at most just outside one standard deviation ($nD_{3/2}$). We therefore suspect differences in the analysis as well as the range of included $n$ values to be the reason for the discrepancy.

The value of the ground state ionization energy (i.e. the binding energy of the ground state), as obtained by fitting equation (\ref{eq:quantendefekt_formel}) to the experimental data, is given in table \ref{tab:IE_Vergleich}. Because all measured Rydberg transition frequencies enter into this value, its statistical error is small (\unit[90]{kHz}) and systematic uncertainties like pressure shifts become significant. We estimate systematic shifts to be on the order of $\unit[200]{kHz}$, resulting in $\unit[\lesssim 300]{kHz}$ uncertainty of the ground state ionization energy. The ground state ionization energy found in this work is within the uncertainty stated by Stoicheff and Weinberger \cite{stoicheff_weinberger} and is two orders of magnitude more accurate.

		\begin{table}[htpb]
		 \begin{tabular}{|l|l||r@.l|r@.l|}
		  \hline
			                      &            & \multicolumn{2}{|c|}{${}^{85}\mathrm{Rb}$ (Li et al \cite{mmwave})}  & \multicolumn{2}{|c|}{${}^{87}\mathrm{Rb}$ (this work)}\\
			\hline
			\hline
			$n S_{1/2}$ & $\delta_0$ &  3&131 180 4(10)   &  3&131 180 7(8)\\
			                      & $\delta_2$ &  0&178 4(6)         &  0&178 7(2)\\
			\hline
			$n D_{3/2}$ & $\delta_0$ &  1&348 091 7(4)  &  1&348 094 8(11)\\
			                      & $\delta_2$ & -0&602 9(3)       & -0&605 4(4)\\
			\hline
			$n D_{5/2}$ & $\delta_0$ &  1&346 465 7(3)   &  1&346 462 2(11)\\
			                      & $\delta_2$ & -0&596 0(2)       & -0&594 0(4)\\
			\hline
		 \end{tabular}
		 \caption{Modified Rydberg-Ritz parameters $\delta_0$ and $\delta_2$ of the $n S$ and $n D$ lines of ${}^{85}\mathrm{Rb}$ following \cite{mmwave} and of ${}^{87}\mathrm{Rb}$ as determined in this work. $n S$ quantum defects refer to the fine structure levels, corrected for shifts due to hyperfine structure.}
		 \label{tab:QD_Vergleich}
		\end{table}

		\begin{table}[htp]
		 \begin{tabular}{|l||c|c|}
			\hline
			                                                          &  Stoicheff and                            & this work\\
																																&    Weinberger \cite{stoicheff_weinberger} & \\
			\hline
			\hline
			$E_{\mathrm{i}\text{,$5 P_{3/2} (F\mathord=3)$}} / h$     & -                & 625.794 214 8(3)\\
																														    &                  & THz\\
			\hline
			$E_{\mathrm{i}\text{,$5 S_{1/2} (F\mathord=1)$}} / (hc)$  & 33 690.945 5(15) & 33 690.946 44(1)\\
																																& $\unit{cm^{-1}}$ & $\unit{cm^{-1}}$\\
			$E_{\mathrm{i}\text{,$5 S_{1/2} (F\mathord=1)$}} / h$   	& 1 010.029 14(5) & 1 010.029 164 6(3)\\
																																& $\unit{THz}$ & $\unit{THz}$\\
			\hline
		 \end{tabular}
		 \caption{Ionization frequency from the $5 P_{3/2} (F\mathord=3)$ state (fit result including systematic error) and the $5 S_{1/2} (F\mathord=1)$ ground state (calculated on basis of the fit result) of ${}^{87}\mathrm{Rb}$ as determined in this work and compared with the value given by Stoicheff and Weinberger \cite{stoicheff_weinberger}.}
		 		 \label{tab:IE_Vergleich}
		\end{table}

\section{conclusions}		

In summary, we performed precision spectroscopy of Rydberg states of ${}^{87}\mathrm{Rb}$ using diode lasers, frequency stabilized to a frequency comb and a calibrated wavelength meter. This allows for absolute frequency measurements with an uncertainty of $\unit[\lesssim 1]{MHz}$.
The system presented in this work combines the ease of use of the wavelength meter with the accuracy of the frequency comb, enabling absolute frequency measurements over a wide frequency range, not limited by the availability of atomic lines to serve as an absolute reference.

We determined quantum defects and the ionization energy of ${}^{87}\mathrm{Rb}$ with an uncertainty of $\unit[\lesssim 300]{kHz}$. In the future, the precision of such measurements can be improved further by reducing the laser linewidths and increasing the bandwidth of the frequency locks.

\section*{acknowledgements}
The authors acknowledge financial support by the BMBF (NanoFutur 03X5506), the European Research Council (ERC Advanced Research Grant ``SOCATHES''), and the Strukturfonds of the University of Tübingen. The authors thank the company HighFinesse for technical support. 
	
\appendix*
\section{Rydberg transition frequencies}

\setlength{\LTcapwidth}{0.45\textwidth}
		
\subsection{$n S_{1/2}$}
\begin{longtable}[t]{|l|r@.l|r@.l|r@.l|}
\hline
\caption{Measured absolute transition frequencies ($\omega_\mathrm{c}$) between the $5 P_{3/2} (F\mathord=3)$ state and $n S_{1/2} (F\mathord=2)$ Rydberg states of ${}^{87}\mathrm{Rb}$. The quantum defects $\delta$ include the correction for hyperfine structure. $\Delta\omega_\mathrm{c}$ is the difference between the measured transition frequency and the frequency calculated with eq.\,(\ref{eq:quantendefekt_formel}) using the quantum defects and ionization energy determined in this work. 
\label{tab:E_n_table_s_1_2}}\\
\endlastfoot
\hline 
$n$ & \multicolumn{2}{|c|}{$\omega_\mathrm{c}$}   & \multicolumn{2}{|c|}{$\delta$} & \multicolumn{2}{|c|}{$\Delta\omega_\mathrm{c}$}\\
    & \multicolumn{2}{|c|}{($\unit{2\pi\ THz}$)}  & \multicolumn{2}{|c|}{}         & \multicolumn{2}{|c|}{($\unit{2\pi\ MHz}$)}\\
 \hline\hline\endhead
 \hline\endfoot
19 & 612&7288381(8) & 3&1318908(5) & -0&8\\
20 & 614&2321542(8) & 3&1318083(6) & 0&5\\
21 & 615&4901687(8) & 3&1317405(7) & -0&2\\
22 & 616&5534925(8) & 3&1316819(8) & 0&7\\
23 & 617&4603214(8) & 3&1316328(10) & 0&5\\
24 & 618&2399264(8) & 3&1315896(11) & 1&0\\
25 & 618&9150382(8) & 3&1315541(13) & 0&2\\
26 & 619&5035327(8) & 3&131521(2) & 0&7\\
27 & 620&0196148(8) & 3&131496(2) & -1&0\\
28 & 620&4746999(8) & 3&131469(2) & 0&3\\
29 & 620&8780318(8) & 3&131448(2) & 0&0\\
30 & 621&2371711(8) & 3&131430(2) & -0&6\\
31 & 621&5583473(8) & 3&131412(3) & -0&4\\
32 & 621&8467280(8) & 3&131395(3) & 0&2\\
33 & 622&1066304(8) & 3&131380(3) & 0&3\\
34 & 622&3416850(8) & 3&131365(4) & 0&8\\
35 & 622&5549595(8) & 3&131358(4) & -0&2\\
36 & 622&7490663(8) & 3&131347(4) & -0&1\\
37 & 622&9262334(8) & 3&131342(5) & -1&0\\
38 & 623&0883784(8) & 3&131330(5) & -0&4\\
39 & 623&2371515(8) & 3&131319(6) & 0&1\\
40 & 623&3739846(8) & 3&131301(6) & 1&4\\
41 & 623&5001176(8) & 3&131314(7) & -1&0\\
42 & 623&6166445(8) & 3&131297(7) & 0&3\\
43 & 623&7245107(8) & 3&131305(8) & -1&2\\
44 & 623&8245598(9) & 3&131277(9) & 1&0\\
45 & 623&9175235(8) & 3&131275(9) & 0&6\\
46 & 624&0040577(8) & 3&131274(10) & 0&3\\
47 & 624&0847416(8) & 3&131276(11) & -0&2\\
48 & 624&1600915(8) & 3&131274(11) & -0&3\\
49 & 624&2305675(8) & 3&131266(12) & -0&0\\
50 & 624&2965811(8) & 3&131252(13) & 0&6\\
51 & 624&3584995(8) & 3&131261(14) & -0&1\\
52 & 624&4166563(8) & 3&131261(14) & -0&3\\
53 & 624&4713499(8) & 3&13126(2) & -0&3\\
54 & 624&5228490(8) & 3&13127(2) & -1&1\\
55 & 624&5714000(8) & 3&13125(2) & -0&0\\
56 & 624&6172212(8) & 3&13124(2) & 0&1\\
57 & 624&6605140(8) & 3&13125(2) & -0&2\\
58 & 624&7014614(8) & 3&13125(2) & -0&4\\
59 & 624&7402301(8) & 3&13124(2) & -0&3\\
60 & 624&7769720(8) & 3&13123(2) & 0&3\\
61 & 624&8118234(8) & 3&13128(2) & -1&6\\
62 & 624&8449163(8) & 3&13126(3) & -0&9\\
63 & 624&8763648(8) & 3&13124(3) & -0&2\\
64 & 624&9062750(8) & 3&13125(3) & -0&5\\
65 & 624&9347470(8) & 3&13125(3) & -0&6
\end{longtable}

\subsection{$n D_{3/2}$}
\begin{longtable}[t]{|l|r@.l|r@.l|r@.l|}
\hline
\caption{Measured Rydberg transition frequencies $5 P_{3/2} (F\mathord=3) \to n D_{3/2}$ of ${}^{87}\mathrm{Rb}$. See description of table \ref{tab:E_n_table_s_1_2}.
}\label{tab:E_n_table_d_3_2}\\
\endlastfoot
\hline 
$n$ & \multicolumn{2}{|c|}{$\omega_\mathrm{c}$}   & \multicolumn{2}{|c|}{$\delta$} & \multicolumn{2}{|c|}{$\Delta\omega_\mathrm{c}$}\\
    & \multicolumn{2}{|c|}{($\unit{2\pi\ THz}$)}  & \multicolumn{2}{|c|}{}         & \multicolumn{2}{|c|}{($\unit{2\pi\ MHz}$)}\\
 \hline\hline\endhead
 \hline\endfoot
19 & 615&2383571(8) & 1&3461520(7) & -0&1\\
20 & 616&3395780(8) & 1&3463543(8) & 0&3\\
21 & 617&2770769(8) & 1&3465269(10) & 0&3\\
22 & 618&0817641(8) & 1&3466765(11) & -0&9\\
23 & 618&7775900(8) & 1&3468033(13) & 0&1\\
24 & 619&3833361(8) & 1&3469144(14) & 0&3\\
25 & 619&9139034(8) & 1&347013(2) & -0&4\\
26 & 620&3812369(8) & 1&347100(2) & -0&4\\
27 & 620&7949991(8) & 1&347174(2) & 0&2\\
28 & 621&1630743(8) & 1&347241(2) & 0&5\\
29 & 621&4919489(8) & 1&347301(3) & 0&6\\
30 & 621&7869975(8) & 1&347356(3) & 0&3\\
31 & 622&0527045(8) & 1&347407(3) & -0&2\\
32 & 622&2928357(8) & 1&347452(4) & -0&3\\
33 & 622&5105718(8) & 1&347488(4) & 0&6\\
34 & 622&7086101(8) & 1&347533(4) & -1&1\\
35 & 622&8892621(8) & 1&347556(5) & 0&8\\
37 & 623&2060305(8) & 1&347618(6) & 0&1\\
38 & 623&3453297(8) & 1&347640(6) & 0&6\\
39 & 623&4736776(8) & 1&347668(7) & -0&0\\
40 & 623&5921942(8) & 1&347689(7) & 0&1\\
41 & 623&7018578(8) & 1&347711(8) & -0&1\\
42 & 623&8035293(8) & 1&347727(8) & 0&1\\
43 & 623&8979667(8) & 1&347740(9) & 0&5\\
44 & 623&9858378(8) & 1&347778(10) & -1&4\\
45 & 624&0677426(8) & 1&347778(10) & -0&1\\
46 & 624&1442056(8) & 1&347789(11) & 0&2\\
47 & 624&2156986(8) & 1&347808(12) & -0&3\\
48 & 624&2826438(8) & 1&347827(13) & -0&7\\
51 & 624&4597844(8) & 1&34784(2) & 0&7\\
53 & 624&5611212(8) & 1&34788(2) & -0&7\\
54 & 624&6075157(8) & 1&34788(2) & -0&2\\
55 & 624&6513410(8) & 1&34786(2) & 0&9\\
57 & 624&7320082(8) & 1&34789(2) & 0&4
\end{longtable}

\subsection{$n D_{5/2}$}
\begin{longtable}[t]{|l|r@.l|r@.l|r@.l|}
\hline
\caption{Measured Rydberg transition frequencies $5 P_{3/2} (F\mathord=3) \to n D_{5/2}$ of ${}^{87}\mathrm{Rb}$. See description of table \ref{tab:E_n_table_s_1_2}.
}\label{tab:E_n_table_d_5_2}\\
\endlastfoot
\hline 
$n$ & \multicolumn{2}{|c|}{$\omega_\mathrm{c}$}   & \multicolumn{2}{|c|}{$\delta$} & \multicolumn{2}{|c|}{$\Delta\omega_\mathrm{c}$}\\
    & \multicolumn{2}{|c|}{($\unit{2\pi\ THz}$)}  & \multicolumn{2}{|c|}{}         & \multicolumn{2}{|c|}{($\unit{2\pi\ MHz}$)}\\
 \hline\hline\endhead
 \hline\endfoot
19 & 615&2402660(8) & 1&3445555(7) & 0&6\\
20 & 616&3411990(8) & 1&3447550(8) & -0&1\\
21 & 617&2784664(8) & 1&3449236(9) & 0&6\\
22 & 618&0829639(8) & 1&3450698(11) & -0&2\\
23 & 618&7786314(8) & 1&3451963(13) & -0&7\\
24 & 619&3842463(8) & 1&3453060(15) & -0&8\\
25 & 619&9147046(8) & 1&345402(2) & -0&7\\
26 & 620&3819464(8) & 1&345484(2) & 0&5\\
27 & 620&7956283(8) & 1&345560(2) & -0&2\\
28 & 621&1636344(8) & 1&345629(2) & -1&1\\
29 & 621&4924518(8) & 1&345685(3) & 0&2\\
30 & 621&7874500(8) & 1&345738(3) & 0&1\\
31 & 622&0531143(8) & 1&345783(3) & 1&0\\
32 & 622&2932066(8) & 1&345828(4) & 0&4\\
33 & 622&5109082(8) & 1&345866(4) & 0&6\\
34 & 622&7089185(8) & 1&345901(4) & 0&8\\
35 & 622&8895425(8) & 1&345931(5) & 1&2\\
37 & 623&2062666(8) & 1&345992(6) & 0&5\\
38 & 623&3455461(8) & 1&346020(6) & -0&0\\
39 & 623&4738779(8) & 1&346042(7) & 0&1\\
40 & 623&5923802(8) & 1&346057(7) & 0&9\\
41 & 623&7020295(8) & 1&346084(8) & 0&0\\
42 & 623&8036883(8) & 1&346104(8) & -0&1\\
43 & 623&8981149(8) & 1&346112(9) & 0&7\\
44 & 623&9859786(8) & 1&346118(10) & 1&5\\
45 & 624&0678704(8) & 1&346162(10) & -0&9\\
46 & 624&1443264(8) & 1&346155(11) & 0&7\\
48 & 624&2827502(8) & 1&346185(13) & 0&3\\
51 & 624&4598714(8) & 1&34622(2) & 0&3\\
53 & 624&5611995(8) & 1&34624(2) & -0&2\\
54 & 624&6075890(8) & 1&34626(2) & -0&4\\
55 & 624&6514104(8) & 1&34623(2) & 0&9\\
57 & 624&7320704(8) & 1&34626(2) & 0&5
\end{longtable}


%

\end{document}